\documentclass[twoside,twocolumn,9pt]{article}
\usepackage{extsizes}
\usepackage[super,sort&compress,comma]{natbib} 
\usepackage[version=3]{mhchem}
\usepackage[left=1.5cm, right=1.5cm, top=1.785cm, bottom=2.0cm]{geometry}
\usepackage{balance}
\usepackage{booktabs}
\usepackage{mathptmx}
\usepackage{sectsty}
\usepackage{graphicx} 
\usepackage{lastpage}
\usepackage{ulem}
\usepackage[format=plain,justification=justified,singlelinecheck=false,font={stretch=1.125,small,sf},labelfont=bf,labelsep=space]{caption}
\usepackage{float}
\usepackage{fancyhdr}
\usepackage{fnpos}
\usepackage[english]{babel}
\addto{\captionsenglish}{%
  \renewcommand{\refname}{Notes and references}
}
\usepackage{array}
\usepackage{droidsans}
\usepackage{charter}
\usepackage[T1]{fontenc}
\usepackage[usenames,dvipsnames]{xcolor}
\usepackage{setspace}
\usepackage[compact]{titlesec}
\usepackage{hyperref}
\usepackage{epstopdf} 
\usepackage{multirow}
\usepackage{dirtytalk}
\usepackage{hyperref}
\usepackage{soul}
\definecolor{cream}{RGB}{222,217,201}

\begin{document}

\pagestyle{fancy}
\thispagestyle{plain}
\fancypagestyle{plain}{
%%%HEADER%%%
\renewcommand{\headrulewidth}{0pt}
}
%%%END OF HEADER%%%
%%% PAGE SETUP 
\makeFNbottom
\makeatletter
\newcommand{\red}[1]{{\color{red} #1}}
\renewcommand\LARGE{\@setfontsize\LARGE{15pt}{17}}
\renewcommand\Large{\@setfontsize\Large{12pt}{14}}
\renewcommand\large{\@setfontsize\large{10pt}{12}}
\renewcommand\footnotesize{\@setfontsize\footnotesize{7pt}{10}}
\makeatother

\renewcommand{\thefootnote}{\fnsymbol{footnote}}
\renewcommand\footnoterule{\vspace*{1pt}% 
\color{cream}\hrule width 3.5in height 0.4pt \color{black}\vspace*{5pt}} 
\setcounter{secnumdepth}{5}

\makeatletter 
\renewcommand\@biblabel[1]{#1}            
\renewcommand\@makefntext[1]% 
{\noindent\makebox[0pt][r]{\@thefnmark\,}#1}
\makeatother 
\renewcommand{\figurename}{\small{Fig.}~}
\sectionfont{\sffamily\Large}
\subsectionfont{\normalsize}
\subsubsectionfont{\bf}
\setstretch{1.125} %In particular, please do not alter this line.
\setlength{\skip\footins}{0.8cm}
\setlength{\footnotesep}{0.25cm}
\setlength{\jot}{10pt}
\titlespacing*{\section}{0pt}{4pt}{4pt}
\titlespacing*{\subsection}{0pt}{15pt}{1pt}
%%%END OF PAGE SETUP%%%

%%%FOOTER%%%
\fancyfoot{}
\fancyfoot[LO,RE]{\vspace{-7.1pt}\includegraphics[height=9pt]{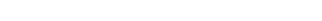}}
\fancyfoot[CO]{\vspace{-7.1pt}\hspace{13.2cm}\includegraphics{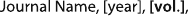}}
\fancyfoot[CE]{\vspace{-7.2pt}\hspace{-14.2cm}\includegraphics{head_foot/RF}}
\fancyfoot[RO]{\footnotesize{\sffamily{1--\pageref{LastPage} ~\textbar  \hspace{2pt}\thepage}}}
\fancyfoot[LE]{\footnotesize{\sffamily{\thepage~\textbar\hspace{3.45cm} 1--\pageref{LastPage}}}}
\fancyhead{}
\renewcommand{\headrulewidth}{0pt} 
\renewcommand{\footrulewidth}{0pt}
\setlength{\arrayrulewidth}{1pt}
\setlength{\columnsep}{6.5mm}
\setlength\bibsep{1pt}
%%%END OF FOOTER%%%

%%%FIGURE SETUP - please do not change any commands within this section%%%
\makeatletter 
\newlength{\figrulesep} 
\setlength{\figrulesep}{0.5\textfloatsep} 

\newcommand{\topfigrule}{\vspace*{-1pt}% 
\noindent{\color{cream}\rule[-\figrulesep]{\columnwidth}{1.5pt}} }

\newcommand{\botfigrule}{\vspace*{-2pt}% 
\noindent{\color{cream}\rule[\figrulesep]{\columnwidth}{1.5pt}} }

\newcommand{\dblfigrule}{\vspace*{-1pt}% 
\noindent{\color{cream}\rule[-\figrulesep]{\textwidth}{1.5pt}} }

\makeatother
%%%END OF FIGURE SETUP%%%

%%%% our commands

\newcommand{\etal}{\textit{et al.\ }}
\newcommand{\maxime}[1]{\textcolor{blue}{#1$|_\mathrm{maxime}$}}
\definecolor{caycecolor}{RGB}{112,91,221}
\definecolor{newcolor}{RGB}{0,176,79}
\newcommand{\cayce}[1]{{\color{caycecolor}#1$|_\mathrm{cayce}$}}
\newcommand{\new}[1]{{\color{newcolor}#1$|_\mathrm{new}$}}
\newcommand{\josh}[1]{{\textcolor{blue}#1$|_\mathrm{josh}$}}

\newcommand{\ie}{{\it i.e., }}
\newcommand{\eg}{{\it e.g. }}

%% Shortcuts

\def\p{\ensuremath\mathbf{p}}
\def\x{\ensuremath\mathbf{x}}
\def\u{\ensuremath\mathbf{u}}
\def\n{\ensuremath\mathbf{n}}

\def\m{\ensuremath\mathbf{m}}
\def\D{\ensuremath\mathbf{D}}
\def\E{\ensuremath\mathbf{E}}
\def\W{\ensuremath\mathbf{W}}
\def\T{\ensuremath\mathbf{T}}
\def\Fc{\ensuremath\mathbf{F}_c}
\def\Fm{\ensuremath\mathbf{F}_m}
\def\FD{\ensuremath\mathbf{F}_D}
\def\Hm{\ensuremath\mathbf{H}_m}
\def\HD{\ensuremath\mathbf{H}_D}
\def\P{\ensuremath{\mathcal{P}}}
\def\Re{\ensuremath{\mathrm{Re}}}
\def\dr{\ensuremath{d_r}}
\def\dt{\ensuremath{d_t}}
\def\sigo{\ensuremath{\sigma_0}}
\def\Vs{\ensuremath{V_s}}

%%%TITLE, AUTHORS AND ABSTRACT%%%
\twocolumn[
  \begin{@twocolumnfalse}
{
\includegraphics[height=30pt]{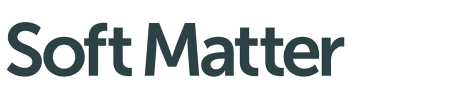}\hfill\raisebox{0pt}[0pt][0pt]{\includegraphics[height=55pt]{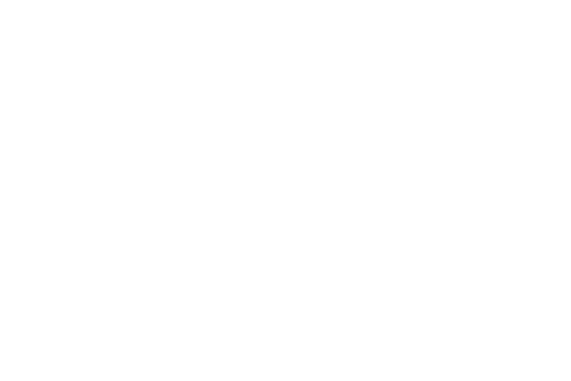}}\\[1ex]
\includegraphics[width=18.5cm]{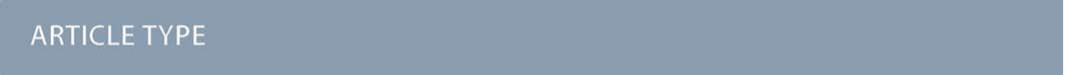}}\par
\vspace{1em}
\sffamily
\begin{tabular}{m{4.5cm} p{13.5cm} }

\includegraphics{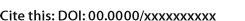} & \noindent\LARGE{\textbf{Kinetically arrested clusters in active filament arrays}} \\

% FINAL TITLE: \includegraphics{head_foot/DOI} & \noindent\LARGE{\textbf{Kinetically arrested self-similar clusters in active filament arrays}} \\
\vspace{0.3cm} & \vspace{0.3cm} \\

 & \noindent\large{Sonu Karayat$^{a}$, Prashant K. Purohit$^{b}$, L Mahadevan$^{c}$, Arvind Gopinath$^{d\ddag}$ and Raghunath Chelakkot$^{a\ddag}$} \\

\includegraphics{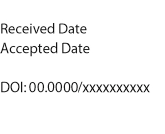} & \noindent\normalsize{
We use Brownian dynamics simulations and theory to study the over-damped spatiotemporal dynamics and pattern formation in a fluid-permeated array of equally spaced, active, elastic filaments that are pinned at one end and free at the other. The filaments are modeled as connected colloidal chains with activity incorporated via compressive follower forces acting along the filament backbone. The length of the chains is smaller than the thermal persistence length. For a range of filament separation and activity values, we find that the filament array eventually self-assembles into a series of regularly spaced, kinetically arrested, compact clusters. Filament activity, geometry, elasticity, and grafting density are each seen to crucially influence the size, shape, and spacing of emergent clusters.  Furthermore, cluster shapes for different grafting densities can be rescaled into self-similar forms with activity-dependent scaling exponents.  We derive theoretical expressions that relate the number of filaments in a cluster and the spacing between clusters, to filament activity, filament elasticity, and grafting density. Our results provide insight into the physical mechanisms involved in the initiation of clustering and suggest that steric contact forces and friction balance active forces and filament elasticity to stabilize the clusters. Our simulations suggest design principles to realize filament-based clusters and similar self-assembling biomimetic materials using active colloids or synthetic microtubule-motor systems.} 
\\

\end{tabular}

 \end{@twocolumnfalse} \vspace{0.6cm}
  ]
%%%FONT SETUP - please do not change any commands within this section
\renewcommand*\rmdefault{bch}\normalfont\upshape
\rmfamily
\section*{}
\vspace{-1cm}

\footnotetext{\textit{$^{a}$~Department of Physics, Indian Institute of Technology Bombay, Mumbai, Maharashtra, India}}
\footnotetext{\textit{$^{b}$~Department of Mechanical Engineering, School of Engineering and Applied Sciences, University of Pennsylvania, Philadelphia, PA, USA}}
\footnotetext{\textit{$^{c}$~School of Engineering and Applied Sciences, Harvard University, Cambridge, MA, USA}}
\footnotetext{\textit{$^{d}$~Department of Bioengineering, University of California Merced,  Merced, CA, USA}}
%Please use \dag to cite the ESI in the main text of the article.
\footnotetext{\ddag~Corresponding authors: agopinath@ucmerced.edu, raghu@phy.iitb.ac.in}
\footnotetext{\dag~Electronic Supplementary Information (ESI) available: [details of any supplementary information available should be included here]. See DOI: 10.1039/cXsm00000x/}

% Use "Eq" instead of "Equation" for equation citations.

\section{Introduction}
Hierarchical structures formed by the assembly of single units or agents are a feature of many biological and synthetic systems. The units themselves may be intrinsically active and possess useful functionality, or they may be passive. In many cases, assembled structures demonstrate unique and sometimes non-trivial emergent properties arising from cooperative effects and assembly-specific interactions~\cite{whitesides2002self}. Apart from the obvious technological relevance of synthesizing hierarchical structures~\cite{whitesides2002self, manoharan2015colloidal}, a detailed investigation of the emergent spatiotemporal properties and functionalities provides insight into the physical mechanisms that underlie these phenomena. For example, studies on self-directed or directed assembly in passive colloidal systems have provided insights into the effects of entropy, interaction potentials and field, and geometric constraints in the formation process~\cite{manoharan2015colloidal}. 

Experimental studies on self-assembly or targeted assembly have employed different types of particles (units); in each case, a variety of physicochemical mechanisms enabling inter-particle interactions have been explored. For example, passive colloids can be driven into organized structures such as chains or clusters by thermal fluctuations, external flows, or by chemical modification such as grafting of nucleotides, polymers, or ligands ~\cite{ong2017programmable, rogers2016using,zhou2024colloidal, mcmullen2022self,cui2021self, liu2020tunable, Nishiguchi2018}. The structural properties of the assemblies can also be controlled by using electric and magnetic fields to modulate or amplify inter-particle interactions~\cite{huang2024colloidal, Nishiguchi2018}. 

Complementing these studies, recent work has focused on active colloids and active particles as structural units to generate multiunit structures. These intrinsically out-of-equilibrium systems include synthetic diffusophoretic colloids, light activated colloids, and living colloidal matter such as bacterial and algal suspensions~\cite{bricard2013emergence, palacci2013living, patteson2016active}. Unlike their passive counterparts, non-equilibrium effects that enable inter-particle interactions are coupled to internal degrees of freedom~\cite{mallory2018active, wang2020active, ramaswamy2010active, Marchetti2013} that may be controlled or adjusted independently without constraints imposed by equilibrium. Aggregates or suspensions of active particles are seen to generate new collective spatiotemporal behaviors including motility-induced phase separation (MIPS)~\cite{fily2015dynamics, Redner2013, Cates2015, digregorioPRL2018}, formation of space-filling and porous clusters, and microphase separation ~\cite{caporusso2020motility, sanoria2024percolation, Das2020, Das2020_pre, kushwaha2024percolation, sanoria2022Percolation}. 

A third class of structural units that show promise as building blocks for the generation of assembled structures are active polymers and filaments. Motivated by the structure and function of biological filamentous structures and assembles such as eukaryotic flagella, cilia and ciliary beds, reconstituted multi-filament active systems containing biofilaments such as actin or microtubules, and molecular motors are being studied {\it in vitro} to create bioinspired materials~\cite{bar2020self, sanchez2011cilia, yadav2024wave}. Recent advances have also been made in designing molecular motors based on biological components with subsequent modifications~\cite{nitta2021printable}. 
Many synthetic systems have also been constructed to mimic the motion of the cilia.
Examples are chains made of microrobots~\cite{xu2024constrained},
elastoactive solids~\cite{zheng2023self}, camphor-infused connected disks~\cite{tiwari2020periodic}, and electrically actuated colloidal chains~\cite{Nishiguchi2018}.

To model the dynamics of multi-filament systems, we need to first understand the response of single active or passive filaments. To aid in theoretical modeling, single colloid chains, biofilament-motor systems, and active polymers in dissipative media are usually treated as slender elastic continuous filaments subject to active and passive forces or torques.  Typically, forces that act along the local tangent vector may be analyzed using follower forces -- these may be localized and applied at the end \cite{de2017spontaneous, fily2020buckling}, or distributed and applied along the filament~\cite{chelakkot2014flagellar, elgeti2015physics, fily2020buckling, sangani2020elastohydrodynamical, Fatehiboroujeni2018, fatehiboroujeni2021three}. The conformational dynamics of active polymers in the limit of moderate to high thermal and non-thermal stochastic noise has also been studied recently\cite{liao2020extensions, winkler2017active, eisenstecken2016conformational}. Fluid-mediated effects on active filaments have also been investigated in detail~\cite{anand2018structure, anand2019behavior, Jayaraman2012, Laskar2013, chakrabarti2019spontaneous}. From these studies, we deduce that the dynamics of a single filament is crucially dependent on the magnitude and distribution of activity, filament elasticity, thermal noise, and drag from the ambient fluid medium. The boundary conditions at the ends of the filament are
an additional crucial feature that plays a role in the selection of dynamical patterns ~\cite{fily2020buckling, fatehiboroujeni2021three}. 

For multi-filament systems, in addition to the parameters discussed above, inter-filament interactions need to be considered. These may be fluid-mediated, long-ranged, and hydrodynamic in origin, or short-ranged surface mediated and steric in origin. The effect of hydrodynamic interactions between filaments, without contact interactions, has been studied in detail and has been shown to cause synchronization and metachronal waves~\cite{elgeti2013emergence,chakrabarti2022multiscale,hickey2023nonreciprocal, von2024hydrodynamic, sangani2020elastohydrodynamical, chakrabarti2019hydrodynamic}. In contrast, even in the absence of a fluid medium, contact interactions in two dimensions have been shown to cause coordination between independently oscillating filaments and sometimes lead to metachronal waves~\cite{zhou2021collective, zhou2022lateral, quillen2021metachronal}. Relaxing the type of boundary constraint, for example, by changing a clamped end to a pinned or pivoted end changes the eventual form of stable multifilament patterns that evolve \cite{chelakkot2021synchronized}. 

Inspired by the ciliary carpets of active filaments seen in the respiratory and reproductive tracts and other bristle-like structures found in biology~\cite{Gopinath2011}, we use Brownian dynamics simulations and theory to study the spatiotemporal dynamics of an array of equally spaced active elastic-anchored filaments. Specifically, we study the emergence of clusters in planar arrays of interacting active filaments anchored at one end. We show that protoclusters initiated by steric interaction and activity coarsen to form kinetically arrested self-similar filament aggregates. These kinetically arrested aggregates are held together by the competition between activity, elasticity, and contact interactions. For a fixed filament length and elasticity, the shape and periodicity of these aggregates can be controlled by changing the inter-filament gap or by varying the filament activity. We identify the important role played by contact interactions here; both normal and tangential contact forces are required to maintain and stabilize these clusters. Rescaling the cluster widths, we find that the cluster shapes can be collapsed into self-similar forms. Guided by these observations, and from further analyses of simulation results for the distribution of forces and torques along filaments in a cluster, we obtain a scaling prediction for the number of filaments within a cluster and an estimate of the cluster size. In general, our simulations and scaling theory provide design principles for the assembly of clusters and similar self-assembled biomimetic materials using chemically modified colloidal beads or synthetic microtubule-motor systems. 

\section{Model and simulation details}

\begin{figure*}[t]
  \centering
  \includegraphics[width=1.7\columnwidth]{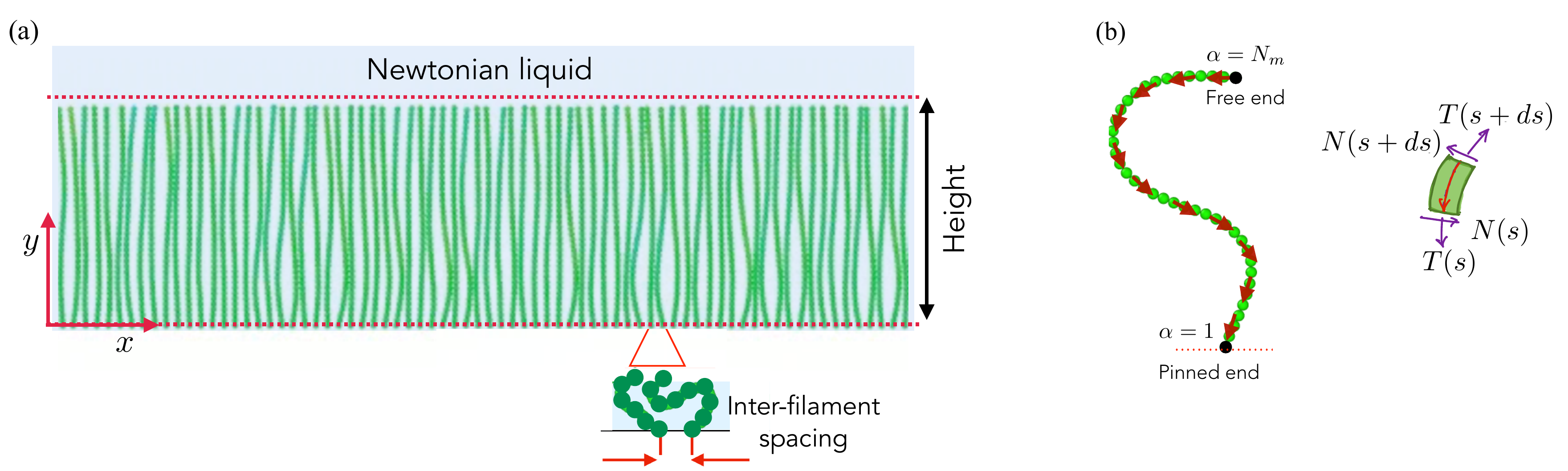}
  \caption{
  {\bf Schematic of the simulation geometry}.
  (a) Shown here is a part of the simulated array of anchored active filaments embedded in a Newtonian viscous liquid. The inter-filament spacing at the anchored based is $\Delta$, the height of the carpet in the non-deformed state is $\ell$. Filaments deform in the two-dimensional $xy$ plane; the transverse $z$ direction is a neutral direction. (b) (left) Snapshot of a single filament illustrating the freely pivoting boundary condition at the base, and the free boundary condition at the top. Active follower forces (red arrows) of constant magnitude act along the tangent and are oriented towards the pinned end. (right) Coarse graining the filament as a continuous elastic curve allows us to study the shape of the filament as resulting from the interplay between activity, restoring elastic forces and torques, and dissipation due to viscous effects. We show here a free-body diagram depicting internal tangential $T$ and normal $N$ forces acting on a small segment of the continuous filament.}
  \label{fig1}
\end{figure*}

In this section, we describe the model and the computational scheme used to analyze the dynamics of the active filament array. The active filament array comprises of $N_{\mathrm{A}}=300$ active filaments (a part of which is shown in Figure~\ref{fig1}(a)), each of length $\ell$, distributed uniformly and aligned in the $x$ direction. The center-to-center spacing between the filaments at the base is $\delta$. All filaments are initially vertically oriented along the $y$ direction. Each filament is a connected chain of $N_{m}= 40$ spherical beads of diameter $\sigma$. The base of each filament is pinned to an anchoring point that allows free rotation about the pinning point (the pivot point), but prevents translation. The end distal to the pinning point is freely suspended and is not subject to external forces or external torques. 

The filament array is permeated by a Newtonian fluid that enables viscous dissipation and generates local drag forces on each bead as it moves. In general, the motion of a bead results in a fluid flow field that influences the motion of beads in neighboring filaments, as well as the motion of more distal areas of the same filament. Our previous work on bead-based discrete filament models \cite{chelakkot2014flagellar} and on related continuum models for analyzing the collective dynamics of single and multiple filaments suggests that local hydrodynamics provides a good physical approximation to results with full hydrodynamics~\cite{sangani2020elastohydrodynamical}. Therefore, we restrict ourselves to the freely draining limit that is often used in bead-spring or bead-rod models of passive polymers. In this limit, each moving bead is subject to a linear isotropic viscous drag force that is antiparallel to the bead velocity. 
Our computational scheme is an adaptation of that used in our previous work on the buckling of single active filaments and clamped filament arrays \cite{chelakkot2014flagellar, fily2020buckling, chelakkot2021synchronized}. Here we summarize the main equations and parameters in the model. 

The beads in each filament are indexed by $\alpha$, and the position vector of the $\alpha ^{\mathrm{th}}$ bead is denoted by ${\bf r}_\alpha$ relative to a stationary global coordinate system. In every filament, each bead, except for the first bead at the anchoring point and the last bead at the free end, is connected to its (two) neighboring beads via linear springs. 
The extensional force between adjacent beads on the same filament due to the springs is derived from the elastic potential $U_{\mathrm{E}}$ 
\begin{equation}
{{U_{\mathrm{E}}} \over k_{\mathrm{B}}T} =\frac{{\kappa_\text{E}} \ell_{0}^{2}}{2 k_{\mathrm{B}}T} \sum_{\alpha=1}^{N_{m}-1} \left( {{|{\bf r}_{\alpha+1} - {\bf r}_\alpha|} \over \ell_{0}} - 1 \right)^2 = \frac{{\kappa_\text{E}} \ell_{0}^{2}}{2 k_{\mathrm{B}}T} \sum_{\alpha=1}^{N_{m}-1} \Phi^{\alpha}_{E}. 
\label{eq1}
\end{equation}

The value of the spring constant $\kappa_{\text{E}}$ is large so that the distance between adjacent beads remains approximately equal to the preferred rest value. The rest length (base state length) of the filament $\ell$, and in our model we choose $\ell_{0}=\sigma$. The center-to-center distance between the first and last beads of the non-deformed straight filament is therefore $(N_{m}-1) \sigma$, and we set this as the filament length $\ell$. The distance from tip to tip spanned by the straight chain is $\ell_{T} = N_{m}\sigma$. In the continuum limit as $N_{m} \rightarrow \infty$ and $\sigma \rightarrow 0$ (that is, $\ell_{0} \rightarrow 0$), $\ell_{T} \rightarrow \ell$, and the filament length may be defined either way.

For a slender filament $\sigma/\ell \ll 1$, bending the filament costs much less energy than stretching it. To model the bending, we implement a three-body bending potential 
\begin{equation}
{{U_{\mathrm{B}}} \over k_{\mathrm{B}}T} = \frac{\kappa \sigma}{2 k_{\mathrm{B}}T}\sum_{\alpha=1}^{N_{m}-1} \Phi_{\mathrm{B}}^{\alpha}, \:\:\:\:\:{\mathrm{where}} \:\:\:\:
\Phi_{\mathrm{B}}^{\alpha}=\left({|{\bf{b}}_{\alpha+1}
-{\bf{b}}_{\alpha}| \over \ell_0}\right)^{2}
\label{eq2}
\end{equation} 
where $\kappa$ is the bending rigidity (with dimensions [Force][Length]$^{2}$), and $\ell_{0} = \sigma$. Equations~\ref{eq1} and \ref{eq2} involve only beads within the {\it same} filament, and the energy thus corresponds to intra-filament energy. Furthermore, 
Eqn. (2) maps to the classical continuum elastic version involving line curvature with the same bending modulus $\kappa$~\footnote{This is confirmed by noting that the line curvature 
${\mathcal{C}} \approx  
|d{\bf b}/ds| \approx |{\bf{b}}_{\alpha+1}
-{\bf{b}}_{\alpha}|/\ell_0
$, ${\bf{b}}_\alpha =({\bf{r}}_{\alpha-1}-{\bf{r}}_{\alpha})/|{\bf{r}}_\alpha-{\bf{r}}_{\alpha-1}|$, and $({\bf b}_{\alpha+1}-{\bf b}_\alpha)/\ell_0\approx d{\bf t}/ds$.}. 

Steric interactions could arise between beads of the same filament, or between beads in neighboring filaments, and thus the expression for the steric interaction potential is nonlocal. Steric interactions between beads in neighboring filaments are implemented via the short-range repulsive WCA (Weeks-Chandler-Anderson) potential. Choosing parameter values such that overlap between {\it neighboring beads} in the same filament does not occur, we define $r_{\alpha \gamma} \equiv |{\bf r}_\alpha-{\bf r}_\gamma|$ as the distance between a pair of spheres ($\alpha,\gamma$) belonging to {\it different filaments}. The steric potential then takes the form
\[
{{U_{\mathrm{WCA}}} \over k_{\mathrm{B}}T} =\frac{\epsilon} {k_{\mathrm{B}}T} \sum_{\alpha=1}^{N_{m}-1} 
%\Phi_{\mathrm{WCA}}^{\alpha}
\sum_{\gamma}
% \neq (\alpha, \alpha -1, \alpha +1)}
    4 
    %\epsilon 
    \left[ \left( \dfrac{\sigma}{r_{\alpha \gamma}} \right)^{12} - \left( \dfrac{\sigma}{r_{\alpha \gamma}} \right)^6 \right] + 1
    \]
\begin{equation}
    =\frac{\epsilon} {k_{\mathrm{B}}T} \sum_{\alpha=1}^{N_{m}-1} 
\Phi_{\mathrm{WCA}}^{\alpha}
     \label{eq_wca}
 \end{equation}
where $r_{\alpha \gamma}<2^{1 \over 6}\sigma$ and $u(r)=0$ otherwise. We extend the definition of the index $\beta$ to include pairs of beads in the same filament as well as in neighboring filaments. Thus, within the same filament, we have $\gamma \neq (\alpha, \alpha -1, \alpha +1)$. 

Inspired by experiments on active Janus beads and colloid bead chains~\footnote 
{An active elastic colloid chain that maps to our computational model may be experimentally realized by elastically connecting surface modified beads.  
An isolated bead when placed in suitable chemical liquid medium self-propels with characteristic speed $v_{0}$ relative to the fluid. The force $f$ acting on a single connected bead can be mapped to the self-propulsion speed of an isolated bead $v_{0}$ in the freely draining limit by $f = v_0/\mu$. Here, $\mu$ is the fluid compliance.}, we make the filaments in our simulation active by making each bead active. This is modeled through follower forces $f$ that act on each bead (Figure~\ref{fig1}(b)); these forces  act along the local tangent vector ${\bf b}_\alpha$. In the continuous limit, this prescription produces a uniform
active force per unit length (force density) $f_{a} = f/\sigma$. 
\begin{table}
\begin{center}
\begin{tabular}{|c|c|c|}
\hline
  Parameter & Interpretation & Value \\
  \hline
   $N_{m}$ &  Number of beads in a filament&  40 \\ 
  %\hline
  $\ell_{0}$ &  Inter-bead distance (set to $\sigma$) &  1 \\ 
  %\hline
  $k_\mathrm{B}T $ & Characteristic energy scale &  1 \\
  %\hline
  $K_\mathrm{E} $ & Extensional modulus &  2 $\times 10^{4}$ \\
  %\hline
  $\epsilon$ & Energy scale in WCA & 1 \\
  %\hline
  $D$ & Translational diffusivity & 1 \\     
  %\hline
  $\mu$ & Mobility (freely-draining limit) & 1 \\  
  %\hline
    $\delta$ & Distance between anchoring points & 2--5 \\  
  %\hline
  $\sigma$ & Range of WCA potential &  1 \\
  %\hline
  $\ell$ & Effective length of each filament &  $39$ \\
  %\hline
  $\kappa$ & Bending rigidity (in $k_{B}T \sigma$ units) &  2 $\times$ $10^{4}$ \\
  \hline
\end{tabular}
\end{center}
\caption{List of simulation parameters, and their scaled values. We set $\ell_{0} = \sigma$ in all our simulations so that microscopically one length scale determines both the steric interaction distance, and the effective surface roughness of each filament.}
\end{table}

Next, we write the equation governing the position ${\bf r}_{\alpha}$ of the bead $\alpha$ in the array. In the over-damped limit, the motion of each bead is driven by the active force, and constrained by the elastic torques due to elastic bending moments, steric interactions due to contact, and stochastic forces due to thermal noise. 
Using $\sigma$, $\sigma^2/D$, and $k_\mathrm{B} T$ as characteristic units of length, time, and energy, we obtain
reduced dimensionless parameters listed in Table 1. Combining equations (1)-(3),  we find that the Langevin equations for bead dynamics (in scaled form, parameter values in Table 1) are 
\begin{equation}
{d{\bf{r}}_{\alpha}  \over dt} = -  \left(\:
\frac{{\kappa_\text{E}}}{2}
{\boldsymbol\nabla} \Phi^{\alpha}_{\text{E}} \: 
+ \:
\frac{\kappa}{2}
{\boldsymbol\nabla} \Phi^{\alpha}_{\kappa} \: + 
{\boldsymbol\nabla} \Phi^{\alpha}_{\text{WCA}}\:\right) + 
f  \:{{\bf b}_\alpha}+\sqrt{2}\:{\mathbf{\zeta}}_\alpha 
\label{eq:EOM}
\end{equation}
where ${\mathbf{\zeta}}_\alpha$ 
is a delta-correlated  noise with zero mean. 

At a coarse-grained level, the nearly inextensible slender colloidal filament may be modeled as a continuous active curve with mean-field elastic properties. Figure 1(c) is a representative free-body diagram that illustrates the forces and torques acting on a small segment of such a filament. 
For fixed filament length, constant bond rest length and bead size, and constant fluid viscosity and temperature, two parameters can be identified in this mean field limit. The first dimensionless parameter quantifies the competition between activity and passive (bending) elasticity, 
\begin{equation}
\beta \equiv {f_{a}\ell^{3} \over \kappa}
%= \left[{1 \over \ell_{0}} \left({v_{0} \over \mu}\right)\right] \left({\ell^{3} \over %\kappa}\right)
= {f \sigma^{2} (N_{\mathrm{m}}-1)^{3} \over {\kappa}}.
%\approx f \sigma^{2}N^{3}_{m}/\kappa
\label{eq:beta-Parameter}
\end{equation}
 We note that
simulations are conducted in scaled units. Thus, to estimate $\beta$ in simulation units, we set $\sigma =1$, the force $f$ acting on the bead is expressed in units of $k_{B}T/\sigma$, and the bending stiffness $\kappa$ is expressed in units of $k_{B}T \sigma$\footnote{
The activity parameter $\beta$ expressed in terms of the tip-to-tip distance is $ f \sigma^{2}N^{3}_{m}/\kappa$.}. The distance between the anchor points of adjacent filaments is $\Delta \equiv \delta/\sigma$.

One end of each filament  ($s=0$) end is pinned along the $x$ axis at regularly spaced intervals $\Delta$. In the simulations, the pinning constraint is implemented through a linear elastic potential (Equation~\ref{eq1}) with restrictions placed on the translational motion of the $s=0$ end monomer. Since we do not employ a bending potential at the fixed end, the filament is free to rotate about the pinning point. The length-scale $\Delta$ determines the density of the array, and controls the strength of steric interactions between neighboring filaments. A dilute array corresponds to large values $\Delta$, while the array becomes increasingly dense as $\Delta \rightarrow 1$. Periodic boundary conditions are imposed along the $x$ direction.
 
As mentioned earlier, we set the distance between two consecutive monomer beads in the filament, $\ell_0$, to be equal to the effective size of the bead (monomer), $(\sigma)$. This limit is relevant to synthetic colloid-based active filament chains and polymers. Thus, the surface of the filament effectively acts as a corrugated surface. As a result, the monomer-monomer interaction imposes effective tangential forces between parallel interacting filaments, causing resistance to a sliding/gliding motion. This tangential resistance is similar to an effective friction force that resists gliding between neighboring filaments or filament groups. As a result of these three effects, the relative movement between filaments can be arrested in favorable cases as adjacently located filaments lock in position with each other.

 The simulations are conducted at a constant temperature.  We introduce a renormalized bending rigidity $\mathcal{K} = \kappa/(N_{m}\sigma k_{B}T)$, which we interpret as the ratio of the thermal persistence length $\kappa/k_{B}T$ to the chain length $N_{m}\sigma$ (here we invoke $N_{m} \gg 1$). Since the simulation parameters ensure that ${\mathcal{K}} \gg 1$, noise alone does not preclude a straight chain, and the shapes of the filaments are predominantly determined by non-thermal forces.  
 
 Each simulation begins with the filaments in their non-deformed state and aligned vertically. The shape of each filament is then evolved by integrating Eq.~\eqref{eq:EOM} in time using an explicit Euler-Maruyama scheme with time step $\Delta t = 10^{-4}$. Typically, steady patterns are obtained with total (scaled) time periods $\sim$ 50. 

\section{Simulation results}

\subsection{Clustering beyond a critical value of $\beta$}

Previous work based on Brownian dynamics simulations, \cite{chelakkot2014flagellar, isele2016dynamics}, and analytical studies~\cite{de2017spontaneous, ling2018instability, Fatehiboroujeni2018, fily2020buckling, sangani2020elastohydrodynamical} indicates that compressive follower forces on a single elastic flexible filament that is clamped at one end and free at the other trigger a transition to periodic oscillatory states provided $\beta$ is sufficiently large. Specifically, when the deformations are restricted to a plane, an isolated clamped-free filament starts to execute flagella-like beats \cite{de2017spontaneous, ling2018instability, Fatehiboroujeni2018, fily2020buckling, sangani2020elastohydrodynamical}. When off-planar deformations are allowed, beyond a critical value of $\beta$, single filaments periodically whirl in three dimensions \cite{ fatehiboroujeni2021three, anand2018structure}.  

Changing the boundary condition at the anchoring point changes the type of instability and the non-linear solutions that emerge. We have earlier shown \cite{chelakkot2014flagellar, fily2020buckling}, that a straight pinned-free filament subject to
tangentially directed compressive forces becomes unstable via a global bifurcation. Post-bifurcation, filaments start to rotate about the pinned end, eventually forming coiled-buckled shapes with well-defined curvatures. Global bifurcation and non-linear solutions persist even in the presence of weak to moderate stochastic (athermal) or thermal noise \cite{chelakkot2014flagellar,  isele2016dynamics, fily2020buckling}, and have also been observed in experiments using connected active colloids \cite{Nishiguchi2018}, and in experiments on motor-driven filament assays \cite{vilfan2019flagella, yadav2024wave}.

When filaments are arrayed in a line and close to each other, the activity-driven buckling instability combines with steric effects and favors the formation of temporary filament aggregates. Specifically,  initial stochastic bending fluctuations of the filament, and subsequent fluctuations in the formation of a coiled state, decide the initial direction of rotation (clockwise or counterclockwise). When two filaments are close together so that $\delta / \ell \ll \beta^{-{1 \over 3}}$, free rotation of either filament at the base is constrained by normal steric interactions. These steric interactions also impact the overall motion of the filaments far from the base, since the follower force is intrinsically coupled to the filament conformation. A further steric effect arises from tangential contact forces that mimic surface friction when two filaments try to slide past each other. This surface effect depends on the ratio $\ell_{0}/\sigma$ and was identified in our earlier work as a parameter controlling spatiotemporal evolution \cite{chelakkot2021synchronized} in active filament arrays with clamped boundary conditions. 
\begin{figure}[h]
  \centering
    \includegraphics[width=0.82\columnwidth]{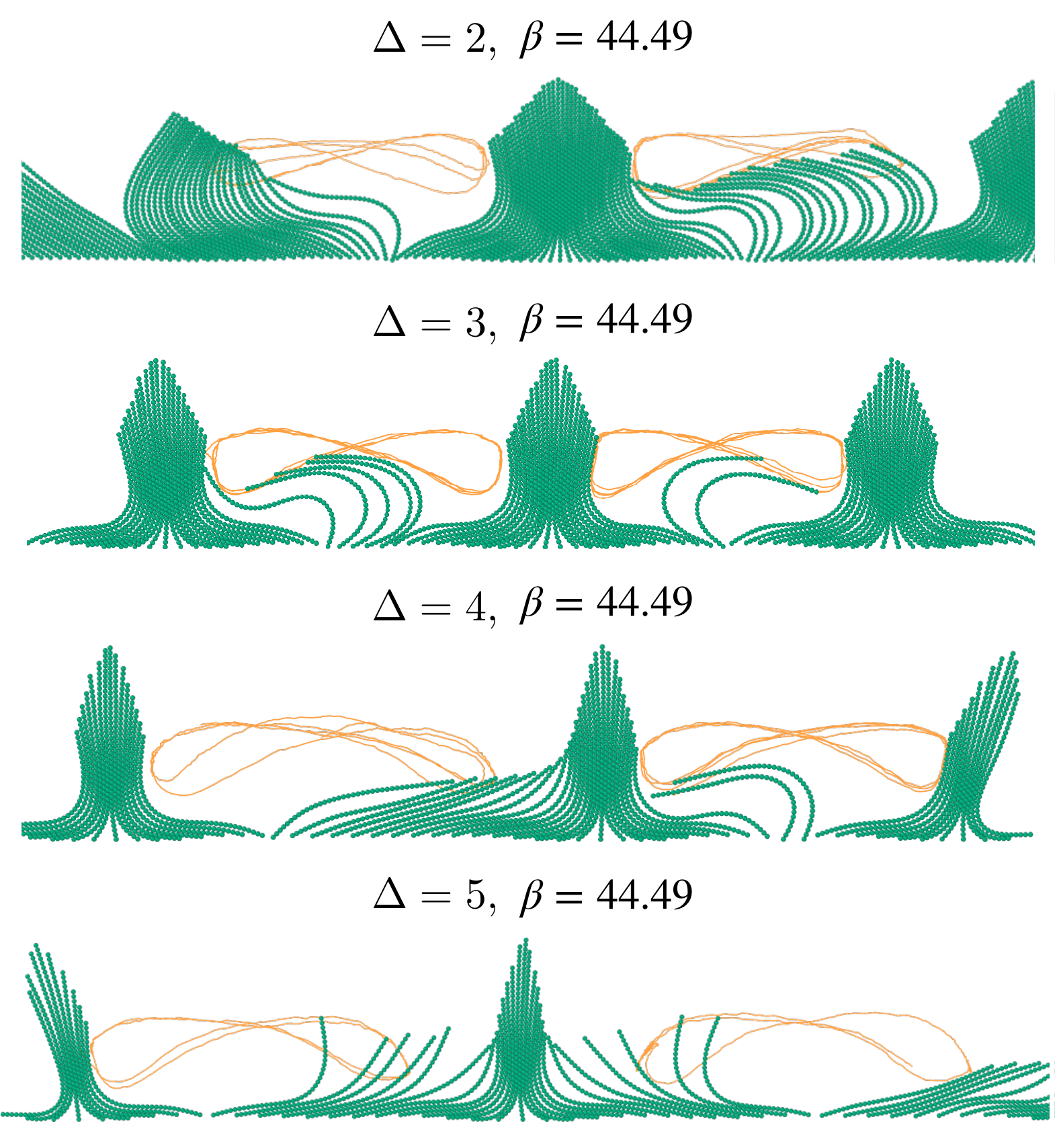}
  \caption{{\bf Kinetically arrested clusters form naturally for different values of the inter-filament spacing}. We show snapshots illustrating clustering when $\Delta$ varies from $2$ to $5$ (top to bottom), and for fixed $\beta = 44.49$. We note how inter-filament spacing controls both the shape of the cluster and the spacing between clusters. Also shown in orange are typical trajectories of the tips of filaments that oscillate {\it between clusters}. The role of changing $\beta$ at constant $\Delta$ is further illustrated in ESM Figure 1.}
  \label{fig2}
\end{figure}
\subsection{Clear clustered states are suppressed for short system sizes}
We tested conditions that result in clustered states by initiating simulations with varying system sizes (number of filaments) and fixed $\Delta$ and $\ell$. We found that for small filament numbers, no clustering is observed, up to the maximum value of $\beta$ investigated ($\sim$ $59.32$). Rather, filaments tilt to one side and fluctuate about this mean angled orientation. For densely packed arrays, collective behavior is dominated by contact forces. For sparse arrays with moderately large values of $\Delta$, significant fluctuations are observed in relative alignment between neighboring filaments.  Since periodic boundary conditions are imposed at the lateral ends of the full array, we deduce that the lack of clustering is related to the length of the simulation domain being too small compared to a naturally preferred inter-cluster wavelength. In fact, by systematically increasing the simulated domain size, we eventually obtain clusters. Increasing the domain length to larger values allows us to obtain kinetically arrested, stable clusters. 

\begin{figure*}[t]
    \centering
 \includegraphics[width=1.6\columnwidth]{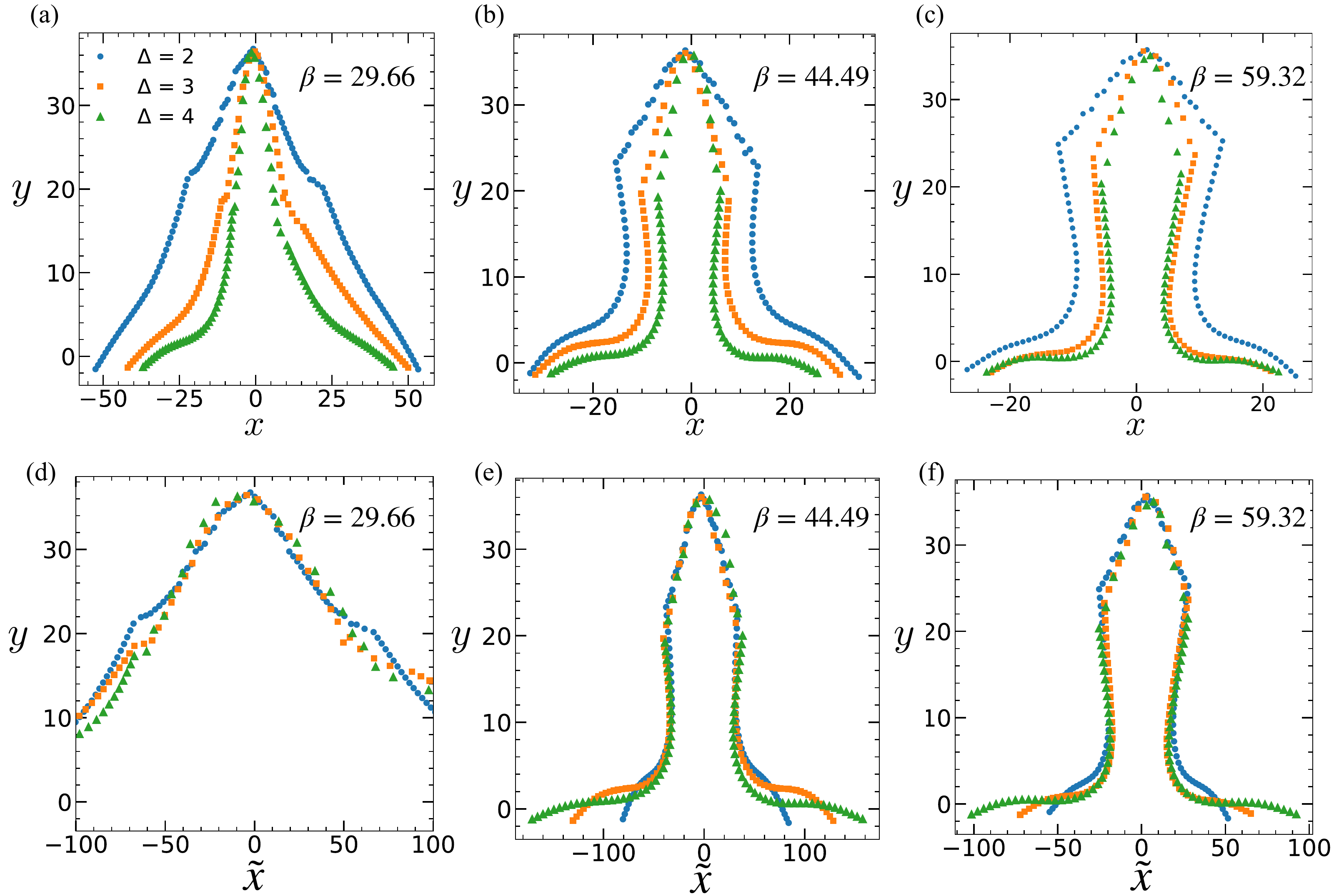}
    \caption{{\bf Outline and collapse of self-similar clusters} for different values of spacing ($\Delta$) and for activity  (a) $\beta = 29.66$ (b) $\beta = 44.49$, and (c) $\beta = 59.32$. %Here we show results for $\beta =29.66$. For ${\Delta = 2}$,  $N_{c} = 53$, the average width of the cluster is $26.60$,  and the maximum width of the cluster is 105.78. For ${\Delta = 3}$, we have $N_{c} = 31$, average cluster width is $22.78$, and the maximum width of the cluster is  $91.95$. For ${\Delta = 4}$, $N_{c} = 21$, the average width of the cluster is $19.83$, and the maximum width of the cluster is  $81.99$. (b) Here we show results for $\beta=44.49$. For ${\Delta = 2}$, we have $N_{c} = 33$, the average width of the cluster is $15.72$, and  the maximum width of the cluster is  $65.76$. For ${\Delta = 3}$,  $N_{c} = 21$, the average width of the cluster is $13.65$, and  the maximum width of the cluster is  $61.97$. For  ${\Delta = 4}$, $N_{c} = 14$, the average width of the cluster is $10.98$, and the maximum width of the cluster is  $54.16$. (c) Here we show cluster shapes for $\beta=59.32$. For  ${\Delta = 2}$, we find that $N_{c} = 26$, the average width of the cluster is $12.40$, and the maximum width of the cluster is  $51.92$. For  ${\Delta = 3}$, $N_{c} = 15$, the average width of the cluster is $9.15$, and the maximum width of the cluster is  $44.09$. For ${\Delta = 4}$, we find $N_{c} = 12$, the average width of the cluster is $8.98$, and the maximum width of the cluster is $46.13$. 
    All length scales (height and width) are in dimensionless units. 
    In sub-figures (d)-(f) we illustrate how scaling allows us to collapse the outlines into self-similar forms. We show the same clusters after scaling the lateral widths along the $x$ axis by a factor $\Delta^{a(\beta)}$, so that $\tilde{x} = x \Delta^{a(\beta)}$. We find that best-fit values of $a$ are (d) $\simeq 1.6$ (e) $\simeq 1.3$  (f) $\simeq 1.05$.}
    \label{fig3}
\end{figure*} 

\subsection{Cluster patterns are controlled by activity and geometry}
\begin{figure*}[h]
\begin{centering}
 \includegraphics[width=1.32\columnwidth]{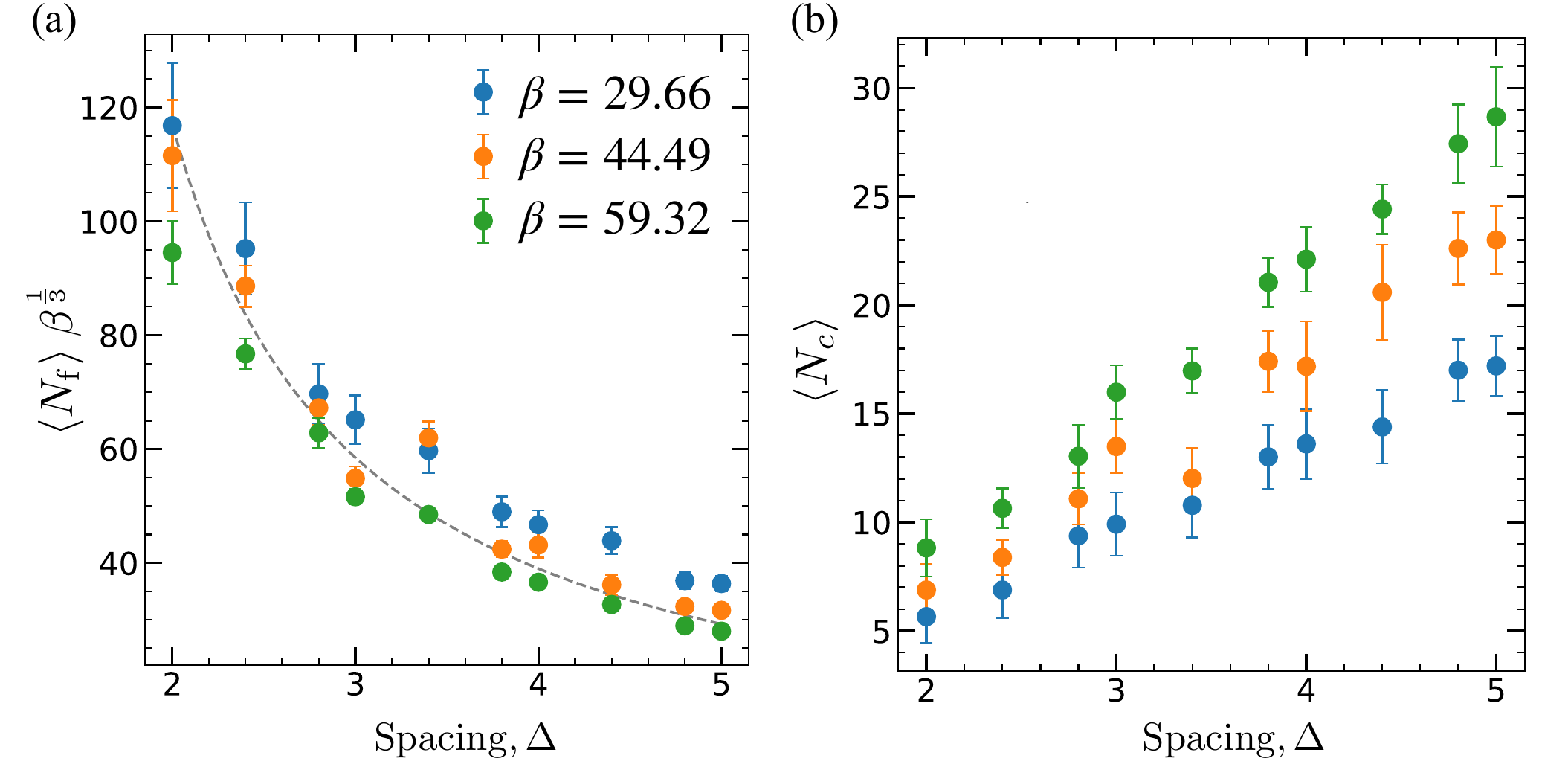}
    \caption{{\bf Cluster properties.} (a) We show how the average number of filaments in a representative cluster, $\langle N_{\mathrm{f}} \rangle $, varies with spacing $\Delta$ for three values of $\beta$. The total number of filaments in the array is fixed at $N_{A}=300$. We note that the size of the domain changes with changing $\Delta$. Inter-cluster oscillating filaments are not taken into account in estimating this cluster size. The dashed line is the prediction from our scaling theory. (b) Here the average number of clusters that form in the array (domain) is shown as a function of $\beta$ and $\Delta$ ($N_{A}=300$). The average number of clusters can be approximated by a linear fit (not shown) with the slopes being $3.8$, $5.5$, and $6.8$ for $\beta=29.66, 44.49$ and $59.32$ respectively. In both (a) and (b), variations in the average value obtained from  simulations are shown as vertical lines. For small $\Delta$, we note large fluctuations in the number of filaments per cluster due to stochastic variations in inter-cluster filaments.}
    \label{fig4}
    \end{centering}
\end{figure*}
\begin{figure*}
     \centering
    \includegraphics[width=1.8\columnwidth]{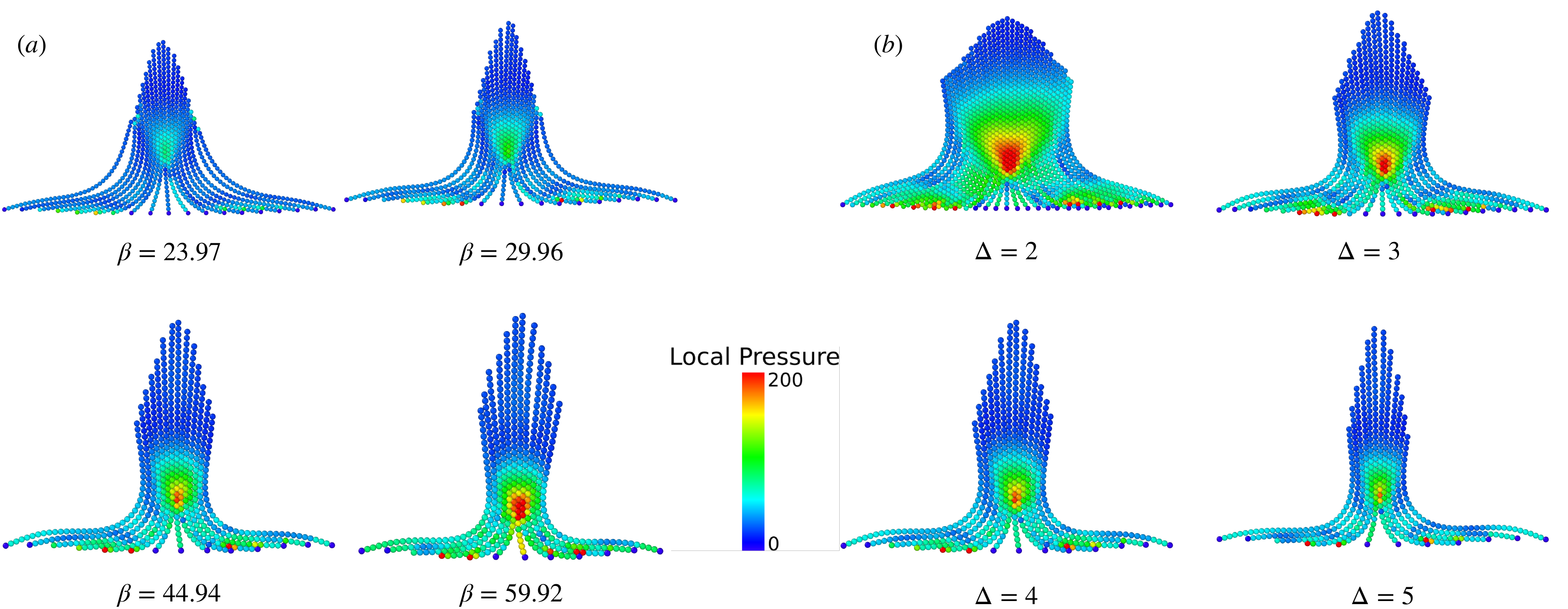}
    \caption{{\bf Spatial distribution of the local pressure inside a cluster}. (a) Here we show the pressure field (averaged over time) in a representative cluster snapshot with simulations where $N = 300$ and $\Delta = 4$ and for different values of $f$. The values of $\beta$ for these plots are $\beta = 23.97, 29.66, 44.49$ and $59.32$. Note that the compact structure corresponding to $\beta = 23.97$ corresponds to a sub-critical value of $\beta$.  
    (b) Time-averaged spatial pressure distributions within a cluster at fixed $N=300$ and $\beta = 44.49$ , while the spacing $\Delta$ varies.}
    \label{fig5}
\end{figure*}

For simulations reported here, we find that the system size $N_{A} = 300$ is sufficiently large to form clearly defined stable clusters provided the activity parameter $\beta \geq 29.66$. In the ESM (see Figures 1(a) and 1(b)), we also present simulation results for a value of $\beta$ lower than the critical value ($\approx 30.6$) for which an isolated pinned-free filament becomes neutrally stable. For this subcritical value of $\beta$, noise (jitter) results in straight non-deformed filaments rotating arbitrarily about the pinned base. As a result of steric interactions and bead-bead locking effects, the filament array is observed to tilt collectively. The tilting motion also triggers isolated regions of filament deformation and transient oscillations, and eventually yields tilted domains interspersed with sparse isolated shapes. As the follower force density increases to $f=10$, the filaments are at the cusp of instability. Here, slow filament rotation, noise, and steric interactions act in concert to yield compact structures.

We next focus on the clustering process for activity parameters that are larger than the critical value for single-filament instability. 
In Fig.~\ref{fig2}, we show the typical shape and configuration of filaments that form kinetically arrested clusters for $\Delta = 2, 3, 4$ and $5$, with the activity parameter $\beta = 44.49$. In ESM Figure 1, we show the typical shape and configuration of the kinetically arrested clusters for varying $\beta$. Visually, it is clear from Fig.~\ref{fig2} and from ESM Figure 1, that the shape and size of the clusters are influenced by both geometry ($\Delta$) and activity ($\beta$). From ESM Figure 1, we deduce that for small $\beta$ with $\Delta$ held constant, the activity-induced deformation of individual filaments is relatively smaller, a typical cluster includes a large number of filaments, and the spacing between clusters is large. However, for larger values of $\beta$, the clusters are relatively compact and are seen to be arrested to unique tower-like shapes (see ESM-Movie1). Typical cluster sizes (in terms of the number of arrested filaments), as well as typical cluster widths, are further strongly impacted by spacing $\Delta$, as evident in Fig.~\ref{fig2}. We find that increasing $\Delta$ decreases the thickness of the cluster for all values of $\beta$ (see ESM-Movie2). We have also observed that the most stable form of the structure tends to be symmetric in shape, as indicated in Fig.~\ref{fig2}. 

Some filaments do not belong to any cluster and are typically found to oscillate between two adjacent clusters. We call these inter-cluster filaments. Such filaments are found to oscillate with a well-defined frequency with low variations when $\Delta$ is small. For larger $\Delta$, the oscillations of inter-cluster filaments can be noisy and erratic. The orientation of such filaments, and especially the tangent vectors at the base, are spatially restricted and constrained due to the steric interactions with the neighboring static filaments belonging to adjacent clusters. Although a single isolated filament moves in a manner consistent with the pinned boundary condition at the base, these inter-cluster filaments show flagella-like oscillations, seen for clamped active filaments~\cite{chelakkot2014flagellar}. Thus, geometric restrictions imposed by arrest cause the pinned filaments to act as if they were effectively clamped and with a shorter length. If the number of such dynamic filaments between two adjacent clusters is large enough, the filaments perform synchronized oscillations.

\subsection{Cluster shapes can be rescaled to self-similar forms}

Although the thickness of the jammed cluster is crucially dependent on $\Delta$, certain characteristics of the tower-shaped shapes appear to be preserved for a given $\beta$ across different $\Delta$. To closely examine these shapes, we plot the positions of periphery monomers of a given cluster, providing the outline of the clusters~\ref{fig3}.  Since cluster shapes are complex, we chose the number of filaments within a cluster as a suitable metric to represent the cluster size. To identify whether a filament was part of the cluster or not, we looked at the shortest distance between filament pairs. If the shortest distance between a pair of filaments $\delta_{ij} < 2^{1/6}\sigma$, where $\sigma$ is the interaction cut-off, we count these filaments as part of the same cluster.  The outermost filament was identified via this criterion. Because of the dynamic filaments that oscillate between some of the clusters, the number of filaments in a cluster, defined using the above criteria, is generally a time-dependent quantity. Since our objective is to understand the formation of kinetically arrested clusters, we ignore these oscillating filaments and consider only those that are static and always part of a given cluster to analyze the cluster shapes. Thus, we obtain a shape profile by tracking the locations of the two end filaments and the free monomers, distal to the pinned end, of the inner filaments. We find that for a fixed value of $\beta$, the width of the cluster increases with a decrease in $\Delta$ as more filaments are dynamically arrested in a single cluster. The vertical height of the cluster remains the same as this is controlled by the central trapped filament, which, to the leading order, retains its non-deformed straight shape. 

Motivated by the nature of these profiles, we next tried to see if re-scaling the axes enabled the collapse of cluster outlines to self-similar or universal shapes. We find that for fixed values of $\beta$,  re-scaling the $x$-axis by $\Delta^{a}$, with $a(\beta) >0$ allowed us to collapse outlines for different values of $\Delta$, as shown in Fig.~\ref{fig3}. Re-scaling worked better for larger values of $\beta$ when more clearly outlined and tightly packed clusters were obtained. We note that the collapse is not perfect, with deviations observed as we approach the pinned ends, as seen in Figures 3(d)-3(f). However, the approximately self-similar shapes suggest that the shape of the kinetically arrested cluster is determined by activity, while its lateral extent or width (or equivalently, the number of filaments in the cluster) is strongly controlled by geometry via the inter-filament spacing parameter $\Delta$.

\subsection{Statistical properties of clusters: size and density}

Next, we analyze how the average size and the density of clusters are influenced by the two control parameters $\beta$ and $\Delta$, for the given system size, $N_{A} = 300$. Since $\Delta$ is a parameter that we vary, the lateral extent of the system also varies. Once the system reaches steady state, both the size and the number of clusters are averaged over time and over ensembles.

Using the criterion described earlier to determine whether a filament belongs to a cluster, we obtain domain averaged values, as well as ensemble averaged values, for the number of filaments with a cluster, $\langle N_{\mathrm{f}} \rangle$. In Fig.~\ref{fig4}(a), we plot the average number of filaments in a cluster as a function of $\Delta$ and for different $\beta$. For fixed $\beta$, the average cluster size is found to decrease monotonically with $\Delta$. We also find that the clusters become more compact with increasing $\beta$. Eventually, the size of the cluster is constrained by the tight hexagonal packing geometry seen near where the lateral width is minimum.  

For the shape of a typical cluster, we also calculated an effective cluster width defined as the average of the local width over its vertical length. For fixed $\Delta$, this quantity is also found to decrease monotonously with $\beta$.  Using this, we estimate the average number of clusters in the domain, $\langle N_{c} \rangle$ and plot this in Fig.~\ref{fig4}(b). We find that $\langle N_{c} \rangle$ is approximately linear in $\Delta$ with slopes depending on $\beta$. 
In both (a) and (b), the variations in the average value obtained from different simulations are shown as vertical lines. Data were obtained from up to five independent runs. The averaging was performed over time and for different initial conditions. The error bars are the standard deviation. 
 \begin{figure}[t]
  \centering
  \includegraphics[width=0.69\columnwidth]{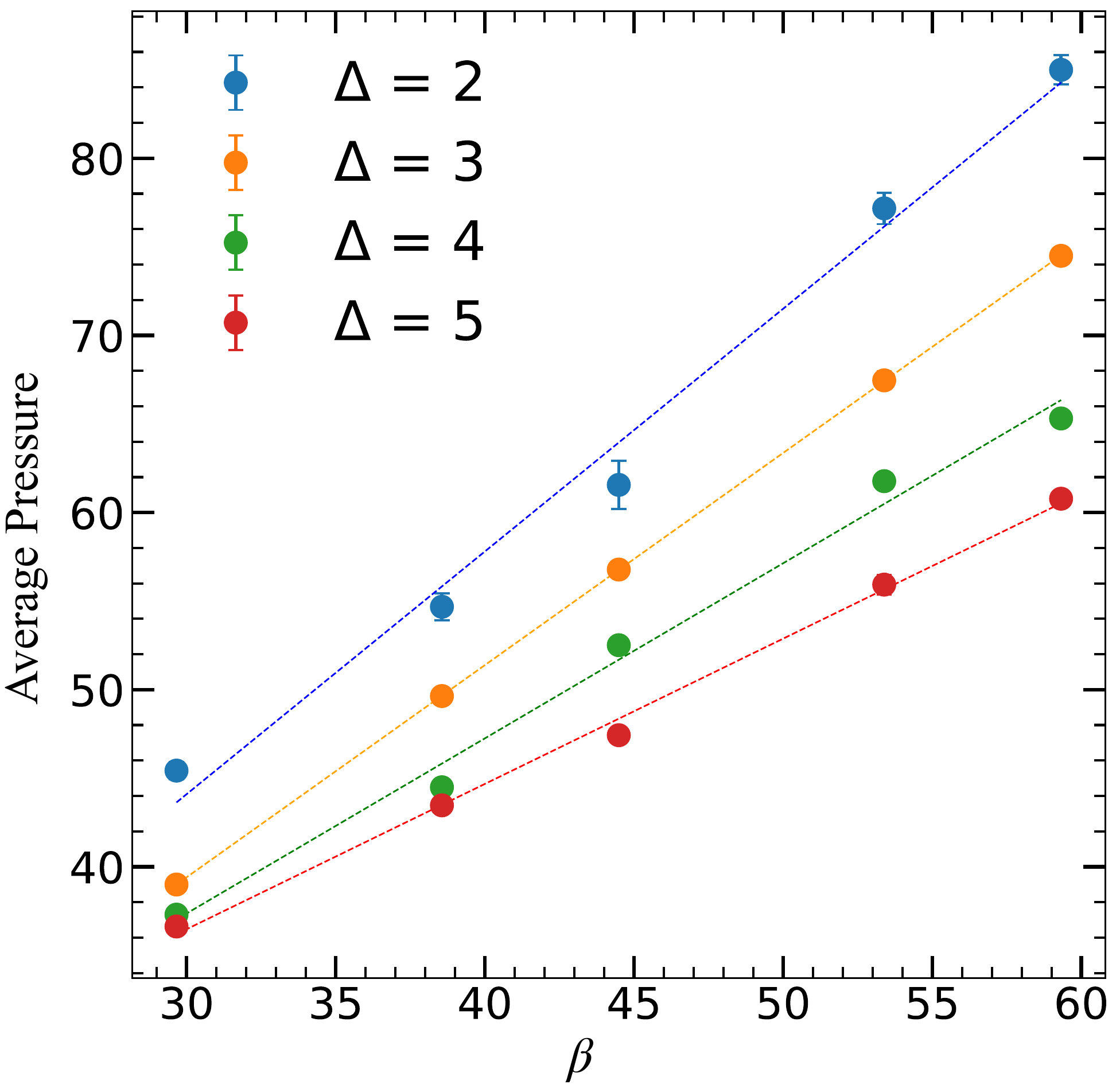}
  \caption{Plot of the (cluster) averaged pressure as a function of the active force density, $\beta$ for various values of $\Delta$. The average pressure here obtained by calculating a combined time and spatial average for each filament in a cluster -- that is, an average over all beads in the cluster. The error bar corresponds to the standard deviation.}
  \label{fig6}
\end{figure}
\begin{figure*}[t]
     \centering
    \includegraphics[width=1.8\columnwidth]{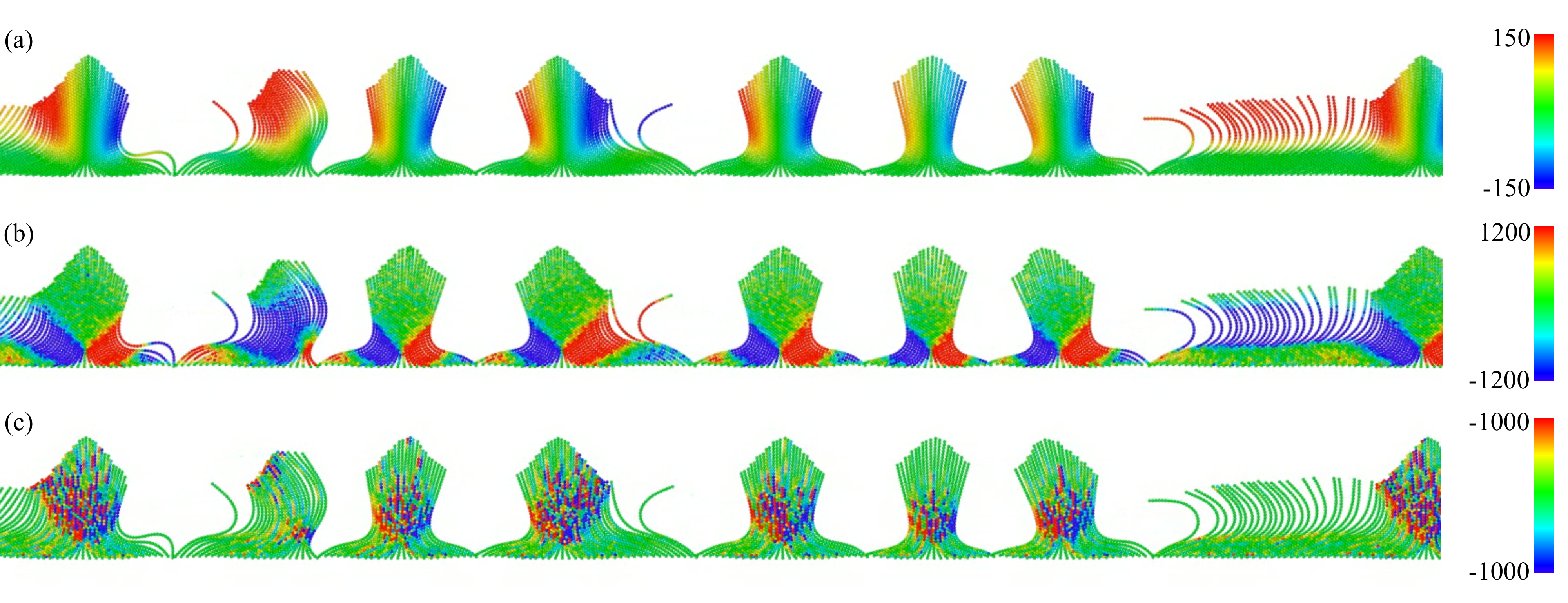}
    \caption{Instantaneous snapshots of the (a) active torque, (b) elastic bending moment, and (c) contact torque on the dynamically arrested clusters and on inter-cluster filaments. The active force density is $\beta = 59.32$,
   %(f=20), 
   and the inter-filament spacing $\Delta=2$.}
    \label{fig7}
\end{figure*}
\subsection{Spatiotemporal variations in internal pressure}

Inspection of the simulation results suggests that cluster stabilization in the most compact part of the cluster is achieved by balancing activity and internal steric stress. To quantify the state of stress within the interior of a cluster, we defined 
a metric quantifying the local interaction stress (pressure), defined for each monomer $\alpha$ and calculated from the pairwise interaction force ${\bf F}_{ \alpha \beta}$ and the separation vector ${\bf r}_{\alpha \beta}$
\begin{equation}
P_{\alpha} = {1 \over N_b} \sum_{\beta \in N_b} {\mathbf{\delta}}:({\bf F}_{  {\alpha \beta}} {\bf r}_{\alpha \beta})
\end{equation}
where $N_b$ is the number of neighboring monomers in contact with the $\alpha^{th}$ monomer, and ${\mathbf{\delta}}$ is the unit tensor. The local interaction pressure was estimated for each monomer in a typical cluster for different $\Delta$ and $\beta$. 

In Fig.~\ref{fig5}, we show the time-averaged value of $P_\alpha$ (color-coded for intensity) superimposed on the respective cluster configuration. As seen in Fig.~\ref{fig5}(a), the local interaction pressure is mostly concentrated within a small region, in the middle of the cluster, where the inter-bead distances are minimum, and the cluster is most tightly packed.  As the activity parameter $\beta$ increases, at fixed $\Delta$, this region moves closer to the basal region, and the local pressure increases in magnitudes. This trend is consistent with the observation that the cluster becomes more compact with increasing $\beta$, as the filaments become more compressed, leading to higher internal pressure. In Fig.~\ref{fig5}(b) we show the local pressure for different $\Delta$, while keeping the activity fixed with $\beta = 44.49$. It is evident that when the cluster thickness is higher for closely packed filaments, the magnitude of local pressure increases. 

We further calculate the global interaction pressure, $P = {1\over N}\sum_{\alpha =1}^N \langle P_\alpha \rangle_t$,
averaging the local, time-averaged, pressure over all 
monomers for different $\Delta$ and $\beta$. 
As shown in Fig.~\ref{fig6}, the value of $P$ increases
approximately linearly with $\beta$ for fixed $\Delta$. 
However, the slope increases as $\Delta$ becomes smaller. 

\section{Discussion}
\begin{figure*}[t]
  \centering
    \includegraphics[width=1.8\columnwidth]{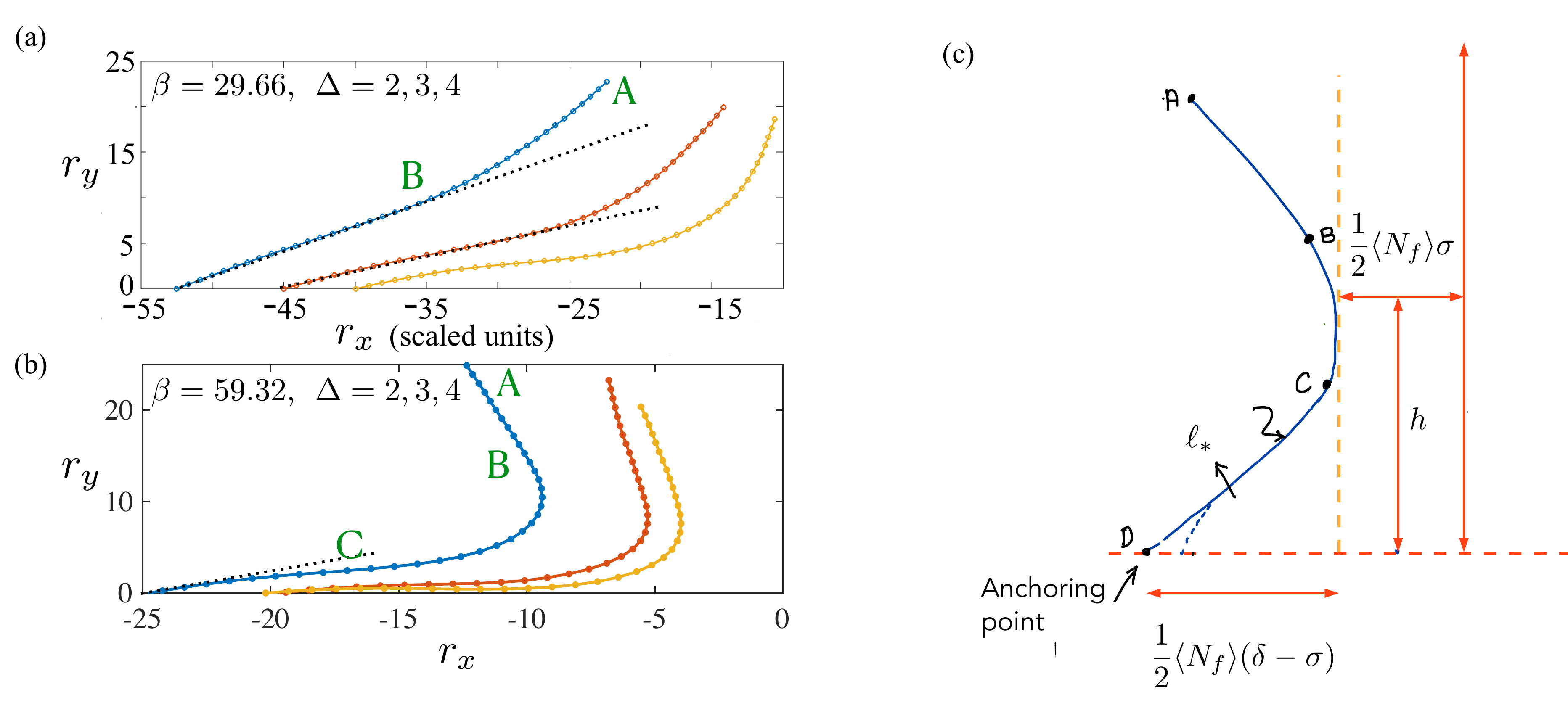}
  \caption{(a,b) Instantaneous filament configuration illustrated via tangent angle plots as a function of colloid location along the filament backbone for the outermost (nearly stationary) filament in a typical cluster. Shapes for two values of $\beta$ are plotted. We show results for (a) $\beta = 29.66$ (when individual non-contacting filaments are close to being neutrally stable), and (b) $\beta = 59.32$ far from instability. For each, the spacing is varied and takes values ${\Delta = 2}$, ${\Delta = 3}$ and ${\Delta = 4}$. The Cartesian components of the beads in the outermost filament --  $r_{x}$ and $r_{y}$ -- are in dimensionless units. (c) Schematic of the outermost filament in the cluster and its location relative to the central filament in the cluster.}
  \label{fig8}
\end{figure*}
\subsection{Initiation of clustering} 
To first understand the emergence of motion leading to the initiation of clustering, compaction, and contact, we consider the linear stability of a single filament aligned vertically and not in contact with any other filaments. We assume that to leading order the follower forces are not sufficiently large to cause filament compression and we therefore assume inextensibility. Detailed analyses of including the nature of bifurcations, critical points, and emergent nonlinear solutions in the absence of steric contact forces have been performed previously by us and others \cite{chelakkot2014flagellar, isele2016dynamics, ling2018instability, fily2020buckling, Fatehiboroujeni2018, sangani2020elastohydrodynamical}. Here we briefly summarize the results that are pertinent to the boundary considered here. 

In ESM Section 1, we derive the equations that govern the tension $T(s)$ and shape $\theta(s)$ of a test filament as a function of the arc-length position $s$ along it. The final equations are
\begin{align}
& 0 = \gamma T'' + \gamma (\theta'' \theta')' - \theta' \:
\left( -\theta''' + T\theta'\right), \\
& 0 = - \theta'''' + (T \theta')' -\dot{\theta}  + \gamma \theta'\:(T'
+ \theta'' \theta' - \beta).
\end{align}
where $\gamma$ is the ratio of the viscous resistances perpendicular and parallel to the filament centerline, and $\dot{\theta}$ is the partial derivative with respect to time of $\theta$. 

To gain insight into the form of the instability, the linear stability of equations (7) and (8) to small perturbations may be analyzed in a simplified setting
where the distributed follower force is replaced by a concentrated point force $f\ell$ acting at the free end~\cite{fily2020buckling}. 
With this approximation for the active force distribution, we rewrite (7) and (8) in terms of a new variable $H$ that is related to $\theta$ by $\theta \approx H'$. The resulting equations are linearized about the base state $H_{0}(s) = 0$ and $T_{0}(s) = {\beta}(s-1)$, and solved with boundary conditions consistent with the pinned-free filament ends 
\begin{equation}
H(0) = H'' (0) = H''(1) = H'''(1) = 0.
\end{equation}
The general solution for this minimal model is 
\begin{equation}
H_{1}(s) = A \: {\mathrm{sinh}} (\lambda_{1}s) + B \:{\mathrm{sin}} (\lambda_{2}s) + C s.
\end{equation}
The solvability condition\cite{fily2020buckling} is found to be
$\lambda_{1} \: {\mathrm{coth}}\:\lambda_{1} = \lambda_{2} \: \mathrm{cot}\:\lambda_{2}$, from which we deduce that
a straight filament is unstable to a global non-oscillatory (rotational) instability via a divergence bifurcation when $\beta \approx 20.2$. 
This value is smaller than the critical value obtained from the linear stability of the exact, approximated equations (Eqns. (7) and (8)), $\beta_{c} \approx 30.6$~\cite{chelakkot2014flagellar, fily2020buckling}. 

Our simulations with $f=10$ yield a patterned array organized into well-formed clusters that are separated by a distance that is $\sim \sigma \Delta$, even though the corresponding activity parameter $\beta = 29.66$ is less than the critical value $\beta_{c} =30.6$ predicted by theory~\cite{fily2020buckling}. This is not surprising since the activity parameter $\beta$ is based on the center-to-center distance, and defining it based on the tip-to-tip distance provides a value $\approx 32$. Furthermore, given the small but nonzero fluctuations in the overall length of the colloid chain, we deduce that $f=10$ ($\beta = 29.66$) is close to the critical point for a single colloid chain to become unstable. 

In the absence of bias and due to intrinsic stochastic noise, the rotation can be clockwise or counterclockwise. Rotating filaments eventually contact their neighbors, whereupon steric interactions and the initial uncoordinated rotation act in concert to generate nascent clusters. These clusters then evolve to their final shape and size.  This coarsening process is also accelerated by oscillating inter-cluster filaments; these push and compact the clusters when they are in contact with them. 

 After instability and far from the critical point, the filaments deform to coiled shapes with characteristic radius $\sim \ell/\beta^{1 \over 3}$. These shapes rotate about the pinning point with a time scale $\tau_{c}$ where $\tau^{-1}_{c}\sim \kappa /(\ell^{4}\zeta_{\|}) \beta^{4 \over 3}$ where $\zeta_{\|}$ is the viscous resistance per unit length for parallel motion~\cite{fily2020buckling}. Thus, we anticipate that filament arrays with spacing $\Delta \gg (\ell / \sigma) \beta^{-{1 \over 3}}$ will not form clusters. When spacing allows for contact and cluster formation, protoclusters form on a time scale $\tau_{c}$. It is important to note that soon after the initial instabilities start, filaments tend to oscillate and behave as clamped-free filaments. This is because steric interactions rapidly become important at the anchored end and effectively provide a stiffening mechanism. 

\subsection{Stabilization of final clustered states} 

The filament dynamics and collective steady-state properties are governed by the active, bending, and interaction forces acting on each filament.  We focus on simulations for fixed $\beta=59.32$ and $\Delta = 2$, for which the clusters can be rescaled to self-similar shapes, as seen in Figure 3 (f). For each we start with $N_{A} = 300$ filaments and follow the development and eventual stabilization of the clustered state. To analyze the roles of the different physical mechanisms in this process, we calculate and track three quantities: local moments generated due to the cumulative active force ${\bf M}^a_\alpha$, moments about the basal end ${\bf M}^i_\alpha$, and local discretized bending moments
${\bf M}_\alpha$. The first two quantities are 
\begin{eqnarray}
{\bf M}^a_\alpha &=& f_a {\bf b}_{\alpha}\times({\bf r}_\alpha - {\bf r}_1) \\
{\bf M}^i_\alpha  &=& \sum_{\beta} {\bf f}^{WCA}_{\alpha \beta}\times({\bf r}_\alpha - {\bf r}_1)
\end{eqnarray}
ESM Movies 1-3 show the evolution of the final clustered patterns starting from the initial vertically aligned filament array. The color scheme in the movies and the intensity of the quantities (in scaled units) correspond to the color map in Fig. \ref{fig7}.

Since the active force acts along the local tangent, the local slope of the filament configuration determines the resultant direction of rotation due to this force (Fig~\ref{fig7}(a)). In general, the direction ${\bf M}^a_\alpha$ can vary along the contour of the filament, depending on the geometry of the filament. A close look at the variation of the bending moment along the filament reveals that the bending moment is predominantly localized in a region where the curvature is large. Interestingly, this region of maximum curvature (bending moment) is located closer to the basal end for the filaments in the interior of the cluster (Fig~\ref{fig7}(b)). However, this region gradually moves towards the distal end of the filament. The curvature field also allows us to identify the region where the local slope changes sign. We find that the direction of the active torque changes across the region of maximum curvature, as is evident from Fig.~\ref{fig7}(a)-(b).  

For a peripheral filament of the cluster, the active torques acting on the segments above the bend region usually rotate the filament towards the cluster, while the active torques acting below the bend region tend to rotate the filament in the opposite direction. As the region of maximum curvature moves towards the distal end, the slope near the basal end changes to enable outward rotation of the filament. The extent of this region increases as a result of which the filament tends to rotate away from the cluster. Thus, the thickness of the cluster is limited by the geometry of the peripheral filament. The interaction torques are important mainly towards the interior of the cluster, where pairwise interactions are more prominent. The role of the resulting interaction torque is to limit the inward rotation of the filaments. 

These observations are consistent with the equations that govern the shape of static filaments in the interior of the cluster (see ESM $\S$A2) that include coarse-grained tangential and normal effective friction coefficients. We deduce from our simulations that these friction coefficients are highly localized to the middle part of the filament and are negligible in the upper part of the filament, as well as in the lower part close to the anchoring point.

\subsection{Scaling theory for the size of clusters} 

We next aim to obtain a scaling theory for the mean number of filaments within a cluster and rationalize the trend seen in Fig.~\ref{fig4}(a).
In Figures 8 (a) and (b), we show the shape of the outermost filament in a cluster for different values of spacing $\Delta$. For small $\beta$ values with activity values close to the critical value, and for small $\delta$ corresponding to a densely packed array, the outermost filament contour can be roughly divided into two dominant regions  AB and BD. This is also suggested in ESM Figure A1(a) for parameter values $\Delta = 2$ and $\beta = 29.66$. The outermost filament contacts its interior neighbor at A, the portion of the filament from A to B is slightly curved, and this region has negligible inter-filament steric interactions. Between B and the anchoring point (the lower part of the filament), steric contact interactions start to again play a role, and contact interactions take values consistent with a stationary filament. Examination of the forces acting on the beads suggests that the compressive forces from the top free segment are roughly balanced by contact forces.  Motion ensues if the lower segment undergoes an active buckling event that causes the segment to bulge outward and away from the enter of the cluster. When this happens, the upper part of the filament slides and eventually peels away from the cluster.

 In the limit of high $\beta$ and large $\Delta$, three prominent regions can be identified as shown in Figure 8(b). The top segment of the filament AB (with A being the distal free end) is characterized by a nearly straight shape with minimal contact interactions. The BC region is a highly curved region where the filament remains pressed against its neighbor. Here, filament beads are tightly packed with significant steric interactions that balance active forces in region BC and the compression of region AB (see ESM, Section A2). In the last CD segment, the filament experiences minimal steric interactions; we observe a slight curvature that is maintained on a characteristic length scale $\ell_{*}$. In a region $O(\sigma\Delta)$ about the anchoring point, steric interactions cause additional variations in the curvature.  

 Guided by these phenomenological observations, we consider a model cluster with the outermost filament located relative to the core of the cluster as shown in  
 Figure 8(c). The geometric center of the cluster is a straight vertically aligned filament (length $\ell$). Let there be $\langle N_{f} \rangle $ filaments in this cluster and let the minimum width of the cluster occur at a vertical height $h$ as shown in the figure.  The half-width of the cluster in the BC region is $\sim {1 \over 2} \langle N_{f} \rangle \sigma$. Although this estimate does not take into account the hexagonally packed nature of the beads, it does account for the increase in the half-width as one moves towards the two ends (B and C). 
 Consider a static outermost filament, and in this filament let the DC segment curve gently over an arc-length $\ell_{*}$. Simulations suggest that the active compression force from segment AB is balanced by steric interactions in BC. Thus the shape of the segment CD is controlled by the active force density and the bending stiffness, and a torque balance delivers  the characteristic curvature of the lower segment CD 
 \begin{equation}
 {\mathcal{R}}^{-1} \sim (f_{a}/\kappa)^{1 \over 3}.
 \end{equation}
When $\sigma/\ell \ll 1$, and for thin slender clusters (as seen when $\beta \gg \beta_{c}$ and $(\Delta-1) \gg 1$), geometrical consistency requires that the following expressions hold true
\begin{eqnarray}
{\mathcal{R}}^{-1} &\sim& {1 \over 2} \langle N_{f} \rangle (\delta - \sigma)/\ell_{*}^{2} \\
\langle N_{f} \rangle &\approx& 2 \ell^{2}_{*}(f_{a}/\kappa)^{1 \over 3} (\delta - \sigma)^{-1}.
\end{eqnarray}
In equations (14) and (15) we have assumed that the number of filaments in a typical cluster is large, $\langle N_{\mathrm{f}} \rangle \gg 1$.
For the outermost filament to be neutrally stable, we further require that $\ell_{*} \sim (\kappa/f_{a})^{1 \over 3}$. The pre-factor depends on the confinement effects and the effective boundary condition. For a dense array, multi-filament steric interactions near the anchoring point D result in an effectively shorter segment CD. Consolidating these results, we obtain the relationship 
\begin{equation}
\langle N_{f} \rangle \beta^{1 \over 3} \sim a^{*}_{1} {\ell \over \sigma} \left({1 \over {\Delta -1}}\right)
\end{equation}
where $a^{*}_{1}$ is roughly constant (a slowly varying function of $\Delta$ and $\beta$). The dashed curve in Figure 4(a) corresponds to 
the equation $\langle N_f \rangle \beta^{1\over 3} \simeq 117 /(\Delta -1)$. Our scaling prediction matches the simulations' results well.

We conclude by estimating the total number of typical clusters, $\langle N_{c} \rangle$ in the simulation domain with $N_{A}$ filaments.
Consider the limits $\beta \gg \beta_{c}$, and $\Delta \geq 2$ so that the inter-filament spacing is large enough for distinct clusters to form. The spacing between clusters scales as the typical lateral amplitude traced by the tip of an oscillating inter-cluster filament. For a filament of length $\ell = (N_{m}-1)\sigma$, the distance swept by an inter-cluster filament, the characteristic oscillation amplitude is
$\sim a^{*}_{2} \ell \beta^{-{1 \over 3}}$.
A simple geometric argument that balances the overall size of the domain with the sum of the distances between cluster-pairs provides a scaling for the mean number of clusters. Let there be $N_{A}$ filaments and let there be $\langle N_{c} \rangle $ clusters in this domain, each with $\langle N_{\mathrm{f}} \rangle$ filaments. We recall that each filament has diameter $\sigma$, and the distance between neighboring filaments is $\delta$. Since we consider a periodic array, the length of the domain is $\sim N_{A}\delta$.
Each cluster spans, at its base, a length $\sim (\langle N_{f} \rangle -1)\delta + \sigma $, and so the total length occupied by the clusters in the domain is $\sim \langle N_{c} \rangle (\langle N_{f} \rangle \delta + \sigma -\delta)
$. We do not consider the closed-packed regime $\Delta \sim 1$, where the inter-cluster gap is $O(a\Delta)$, because well-separated clusters do not form in this limit. Focusing on well-separated clusters, the gap between each cluster-pair $\sim a^{*}_{2} \ell \beta^{-{1 \over 3}}$. 
Invoking the periodicity of the array again, we find that the sum of these inter-cluster gaps in the domain add up to a length $\sim \langle N_{c} \rangle a^{*}_{2} \ell \beta^{-{1 \over 3}}$.
Balancing lengths then provides the relationship
%\begin{equation}
%    \left(\langle N_{f} \rangle \Delta + (1 - \Delta) \right) \langle N_{c} %\rangle +  a^{*}_{2} \langle N_{c} \rangle (\ell/\sigma) \beta^{-{1 \over 3}} \sim %N_{A} \Delta 
%\end{equation}
$
    \langle N_{c} \rangle \left[\langle N_{f} \rangle \Delta + (1 - \Delta) +  a^{*}_{2} (\ell/\sigma) \beta^{-{1 \over 3}}\right] \sim N_{A} \Delta 
$
which may be written as 
\begin{equation}
    \langle N_{c} \rangle \left[\langle N_{f} \rangle \beta^{1 \over 3} \Delta + (1 - \Delta)\beta^{1 \over 3} +  a^{*}_{2} (\ell/\sigma)\right] \sim N_{A} \Delta \beta^{1 \over 3}
\end{equation}
Substituting Eqn. (16) in Eqn. (17) we get 
\begin{equation}
     \langle N_{c} \rangle \sim N_{A} \Delta \beta^{1 \over 3} \left[ {\ell \over \sigma}  \left(a_{1}^{*} {\Delta \over \Delta -1} + a_{2}^{*} \right) - (\Delta - 1)\beta^{1 \over 3} \right]^{-1}
\end{equation}
Equations (16) and (18) provide asymptotic scaling relationships for the effective cluster width and the effective density of clusters. In deriving these, we ignored variations of $a^{*}_{1}$ and $a^{*}_{2}$ with $\Delta$ and $\beta$, and assumed that we are far from the critical point below which clusters do not form.
We cannot compare Eqn.(18) directly with our simulation results, since $a_{1}^{*}$ and $a_{2}^{*}$ are unknown. However, some observations can be made. In our simulations, $\ell/\sigma \sim 40$, and the maximum value of $(\Delta-1)\beta^{1 \over 3}$ is $\sim 15.6$. If the first term in the square brackets were much larger than the second, which is reasonable since we expect $a_{1}^{*}$ and $a_{2}^{*}$ to both be $\sim  O(1)$ or larger, then at fixed $\Delta$, $\langle N_{c} \rangle \sim \beta^{1 \over 3}$. Similarly, at fixed $\beta$ and when moderate to large $\Delta$, we expect $\langle N_{c} \rangle \sim \Delta$. Both of these trends are consistent with Figure 4(c).

\section{Conclusions}
Current theoretical work on active chains, filaments, and polymers focuses on understanding the dynamics of single filaments with simplified~\cite{chelakkot2014flagellar, chelakkot2021synchronized, de2017spontaneous, eisenstecken2016conformational, fily2020buckling, Fatehiboroujeni2018, fatehiboroujeni2021three, liao2020extensions, ling2018instability},  or detailed hydrodynamics~\cite{chakrabarti2019spontaneous, laskar2017filament, sangani2020elastohydrodynamical}. The work presented here on multi-filament arrays complements and extends these previous studies.
We show that steric interactions between filaments and associated effective frictional and locking effects combine with
activity-driven linear instabilities to yield highly non-equilibrium kinetically arrested self-similar clusters.  In general, our simulations and theory provide design principles for assembling clusters and other self-assembled biomimetic materials. Our simulations show that the shape and spacing of the clusters can be controlled by varying geometry (in dimensional terms, the filament length $\ell$, the inter-filament spacing $\Delta$, and the filament thickness $\sigma$), by changing activity (by changing the force per unit length $f_{\mathrm{a}}$), and bending elasticity (via the dimensional bending stiffness $B$).  

We conclude with some comments on how the multi-filament arrays studied here may be realized in experimental settings at the microfluidic scale. The first route is to use reconstituted filament-motor assays or synthetic mimics of these as the activity can be controlled by changing the motor density or ATP concentration. In previous work, we estimated \cite{fily2020buckling} that
for microtubule-kinesin systems  $\ell \sim 5-30 \:\mu$m, and $\beta \sim 10^{4} - 10^{5}$. For reconstituted actin-myosin assemblies, the values $\ell \sim 1-4\ \mu$m are typical, with $\beta \sim 10^2 - 10^{3}$. To obtain lower values of $\beta$, the density of the motors can be reduced or the number of motor attachment sites can be adjusted. 
The second route is to use active colloids, such as self-propelling, diffusophoretic, or field-activated colloidal beads connected in a chain-like configuration. Flexible colloidal chains of lengths $ \ell \sim O(10)$ $\mu$m formed by connecting beads using polymeric biomolecular linkers such as biotin-streptavadin bonds, electrostatic interactions, or induced dipolar interactions are structurally stable and can be constructed with flexible bending moduli (as demonstrated in~\cite{Nishiguchi2018}). 
Individual active colloidal chains or polymers may then be grafted on to a surface using linkers that permit torsional flexibility at the grafted end, thus mimicking the pinned boundary condition. Recent work on polymeric pillar-based canopies and pH-dependent frictional systems~\cite{chau2023ph} suggests extensions of our results to synthesizing clusters in the millimeter range. 

Experimentally, arrayed filament clusters as described in this work may be realized in Hele-Shaw geometries, where the displacements and deformations of the anchored filaments can be confined to a thin plane. In future research, we will relax this assumption and study spatiotemporal patterns and clustering that form when filaments arrayed on a two-dimensional surface. Previous theoretical studies suggest that isolated filaments may undergo out-of-plane instabilities that result in twirling~\cite{fatehiboroujeni2021three} or rotating states~\cite{ling2018instability}. These studies suggest that active filaments anchored periodically on a surface will undergo three-dimensional motions which, when combined with steric interactions and roughness-based locking at the colloid scale, may yield compact cone-like filament clusters or twisted bundles. Structures with such geometries are seen in biology and many serve as mechanosensing structures. An example is the inner-ear hair cell mechanotransduction pathway where the deflection of hair bundles, the sensory organelles of hair cells, activates mechanically-gated channels (MCGs)~\cite{gillespie2009mechanotransduction}. Future work related to the design of these bioinspired structures is an intriguing and exciting prospect. 

\section*{Acknowledgements}
AG acknowledges funding and support from the National Science Foundation (NSF) through the NSF CAREER award 2047210 and also the NSF award 2026782.
\section*{Author contributions statement}
AG and RC conceptualized the problem and formulated the model. RC and SK wrote the code and performed the simulations. AG developed the mean-field model. AG, PKP and LM developed the scaling theory. AG and RC wrote the original manuscript draft. AG, RC, PKP and LM finalized the manuscript. All authors read and contributed to the final version of the manuscript.
\section*{Additional information}
\textbf{Competing interests} 
The authors declare no competing interests.
\balance
\renewcommand\refname{References}
\bibliography{Citation-Cluster}
\bibliographystyle{rsc}

\end{document}